# Quelle architecture implémenter pour intégrer des données du monde réel, dans un simulateur Unreal Engine 5 et dans le contexte de la réalité mixte ?

Jonathan CASSAING
*Extrait thèse professionnelle, Mastère Spécialisé Systèmes Informatiques Ouverts
CentraleSupélec, Université Paris Saclay, France*

(Abstract) Due to its ability to generate millions of particles, massively detailed scenes and confusing artificial illumination with reality, the version 5 of Unreal Engine promises unprecedented industrial applications.

The paradigms and aims of Unreal Engine contrast with the industrial simulators typically used by the scientific community. The visual quality and performance of its rendering engine increase the opportunities, especially for industries and simulation business: where interoperability and scalability are required.

The study of the following issue « Which architecture should we implement to integrate real-world data, in an Unreal Engine 5 simulator and in a mixed-reality environment? » offers a point of view. The topic is reexamined in an innovative and conceptual way, such as the generalization of mixed-reality technologies, Internet of Things, digital twins, Big Data but providing a solution for simple and actual use cases.

This paper gives a detailed analysis of the issue, at both theoretical and operational level. Then, the document goes deep into Unreal Engine's operation in order to extract the vanilla capabilities. Next, the C++ Plugin system is reviewed in details as well as the third-party library integration: pitfalls to be avoided are shown. Finally, the last chapter proposes a generic architecture, useful in large-scale industrial 3D applications, such as collaborative work or hyper-connected simulators.

This document might be of interest to an Unreal Engine expert who would like to discover about server architectures. Conversely, it could be relevant for an expert in backend servers who wants to learn about Unreal Engine capabilities.

This research concludes that Unreal Engine's modularity enables integration with almost any protocol. The features to integrate external real data are numerous but depend on use cases. Distributed systems for Big Data require a scalable architecture, possibly without the use of the Unreal Engine dedicated server. Environments, which require sub-second latency need to implement direct connections, bypassing any intermediate servers.

# Table des matières

# 1. LES OBJECTIFS DE L'ÉTUDE

Analyser le sujet, « Quelle architecture implémenter pour intégrer des données du monde réel, dans un simulateur Unreal Engine 5 (UE) et dans le contexte de la réalité mixte ? », nécessite de le confronter avec l'utilité et les objectifs liés aux simulateurs immersifs ou aux mondes virtuels augmentés. Impossible de se passer d'une comparaison avec le concept de métaverse qui semble correspondre à un cas d'usage très générique et englobant (une définition du terme métaverse est proposée au titre 7). Sans spéculer sur ce concept encore théorique à ce jour et en s'écartant de l'effet de mode qu'il produit, le concept de métaverse partage une multitude de cas d'utilisations des technologies de réalité étendue (XR – miXed Reality), et qui peuvent exister individuellement. Même si le concept de métaverse partage les mêmes préoccupations que de nombreuses applications XR indépendantes, il doit être considéré comme une discipline différente [1].

Voici un ensemble non exhaustif d'applications XR concrètes et plus actuelles : l'entrainement et la formation (l'éducation, la santé, l'armée, l'industrie), l'industrie (maintenance, production), les jeux vidéo (Pokemon Go) et événement artistiques (concerts), les réseaux sociaux (Rec Room, Microsoft AltspaceVR), le travail collaboratif, les conférences et évènements, les jumeaux numériques, la modélisation 3D (architecture, conception, design intérieurs), le tourisme, etc.

Le besoin d'interconnexion pour la majorité des cas d'usages cités est fort. Qu'il s'agisse de se baser sur une technologie propriétaire ou sur un standard ouvert, il est souvent question d'interopérabilité entre le système XR, l'environnement virtuel et un système informatique qui fournit un service ou qui déploie une donnée.

## 1.1. Des normes, standards ou protocoles ouverts

Le besoin d'interopérabilité étant élevé dans la majorité des cas d'usage des technologies de réalité mixtes, une réponse au sujet doit pouvoir intéresser un maximum de ces applications. Il en découle qu'il ne sera pas étudié dans ce document les solutions reposantes sur des protocoles ou technologies propriétaires et fermées. Le choix de protocoles qui s'appuient sur des normes ou standards est donc primordial. S'agissant de s'interfacer avec Unreal Engine, moteur pouvant fonctionner sous Microsoft Windows, Linux ou macOS, la présente étude ne s'intéresse pas aux protocoles des couches matérielles du modèle OSI (Open Systems Interconnection) [2]. Quant à la couche de transport, cette dernière n'a pas de raison de différer des standards UDP (User Datagram Protocol) [3] et TCP (Transmission Control Protocol) [4]. Les couches hautes restantes doivent s'appuyer sur un maximum de normes, standards ou protocoles ouverts.

Par nature, Unreal Engine est une application dite « client lourd » : c'est-à-dire que l'utilisateur final doit installer le simulateur pour profiter de l'application développée. Par opposition, les applications dites « client léger » s'appuient sur des logiciels existants et souvent déjà installées : comme un navigateur web. Cette distinction peut avoir un impact sur le cas d'utilisation final où, le choix d'un client léger serait plus pertinent. Cependant, dans la majorité des cas Unreal Engine reste pertinent parce qu'il offre un niveau de performance plus élevée qu'une application conçue dans un navigateur : et la réalité virtuelle consomme beaucoup de ressources. Le choix entre Unreal Engine et un client léger n'est pas l'objet du document mais il est essentiel que la question soit évaluée lors du choix des composants.

> Il est à noter que Unreal Engine possède des capacités « client léger ». En effet, il est possible d'embarquer une application développée, sous Unreal Engine, dans un navigateur. Ces capacités n'ont pas été évaluées, ni leur intégration avec un casque de réalité virtuelle. Epic Games, l'éditeur d'Unreal Engine, indique que le support des navigateurs a été migré, depuis le moteur, vers leur plateforme d'extensions publiques. La documentations de cette partie n'est plus hébergée par Epic Games mais sur GitHub.com.

### 1.1.1. De la performance

La performance des solutions d'architectures retenues constitue le critère numéro 2. S'agissant de disposer en bout de chaine, d'un moteur 3D capable de générer de la réalité virtuelle, la performance compte toujours. Cependant, il n'est pas inutile d'être plus précis. En effet, en réalité le niveau de performance nécessaire dépend : du cas d'usage mais aussi des impératifs et caractéristiques des données à faire transiter entre Unreal Engine et le reste du système d'information ou de la chaine de traitement. Les données, selon leur format, nature ou caractéristique, vont nécessiter d'adapter l'architecture et les protocoles pour que le besoin soit satisfait.

Les exigences de latence font partie des plus difficiles à satisfaire. La latence, est le temps qui s'écoule entre la production d'une donnée et sa réception (ou sa visualisation) : c'est-à-dire qu'elle porte le temps passé à travers le réseau.

Dans les systèmes les plus extrêmes, il est nécessaire d'employer des réseaux dédiés, des systèmes temps réels ou des cartes électroniques conçues spécifiquement (à base de FPGA « Field-Programmable Gate Array » par exemple). Le moteur Unreal Engine n'est pas un moteur temps réel au sens de « l'informatique temps réel » : par conséquent, de telles exigences risquent d'être peine perdue à satisfaire.

> Avant d'aller plus loin sur la notion de temps réel, il est nécessaire de la définir car elle trouve plusieurs définitions selon le contexte où elle est employée, en informatique.
>
> En informatique embarquée (informatique temps réel), un système est temps réel lorsque ce dernier doit satisfaire des contraintes de temps, qui sont fermes, en plus de celles de délivrer les bons résultats. Concrètement, on doit être capable de maîtriser le temps exact qu'il va être nécessaire au système pour que ce dernier exécute un ordre/une tâche. Exemple : un système d'Airbag est un système temps réel (trop tôt ou trop tard ne sont pas des cas acceptables, ordre de grandeur de 15 à 50 millisecondes suivant les systèmes). Il y a une notion de maîtrise du temps, mais pas nécessairement de vitesse [5]. En informatique, ces systèmes nécessitent la mise en œuvre de systèmes d'exploitation temps réel, qui sont bien loin de notre Debian, Windows 11 ou macOS Sierra (pour n'oublier personne).
>
> Dans l'univers des moteurs de jeux vidéo, ces derniers sont également indiqués « temps réel ». Ici, cette notion est plus subjective et fait appel à la fréquence à laquelle le moteur est capable de générer des images. Cette fréquence doit rester suffisamment élevée pour que le rendu paraisse fluide pour l'œil humain. La tolérance de temps est donc plus floue, voir différente d'une personne à une autre. Les films emploient généralement une fréquence de 24 IPS (Images Par Secondes), soit un délai de 41,7 ms entre chaque image. Les moteurs de jeu vidéo montent beaucoup plus haut, ce qui améliore l'agrément. Il est courant de voir des simulateurs fonctionner à 144 IPS, soit 6,9 ms entre chaque image. Notez que le moteur peut être plus véloce que notre système d'Airbag, il n'est cependant pas réellement temps réel au sens défini par l'univers de l'embarqué : car il n'y a aucune garantie que les images rendues le seront toutes les 6,9 ms.
>
> Dans l'univers des systèmes d'information, il est commun de parler de donnée temps réel. Par exemple, il peut s'agir de récupérer « la météorologie en temps réel ». La notion de temps réel fait référence à celle de « temps utile » : c'est-à-dire la fraicheur de la donnée qui doit être compatible avec son utilité. Actualiser la météo toutes les heures pourrait par exemple suffire pour certaines applications.

Concernant Unreal Engine, pour que ce dernier reste « temps réel », l'architecture et les protocoles retenus ne doivent pas impacter trop la fréquence d'images rendues. On peut communément considérer que le moteur n'est plus vraiment temps réel si la fréquence d'image descend sous les 24 IPS (41.7 ms).

Certaines applications exigent des contraintes de faible latence : par exemple, piloter un drone depuis un casque de réalité virtuelle, dans un monde simulé immersif, qui reçoit le flux d'images filmées par le drone et qui envoie à ce dernier des ordres de contrôles/commandes (sur ces axes de tangage, roulis et lacet pour en contrôler la rotation).

Enfin, les caractéristiques de vélocité et de volume des données à recevoir/envoyer dans/depuis Unreal Engine peuvent impacter significativement les performances.

### 1.1.2. Une facilité d'intégration

Avant d'évaluer la facilité d'intégration d'une solution, avec Unreal Engine, il faut en évaluer sa faisabilité. Unreal Engine est multiplateforme : Microsoft Windows, Linux et Apple macOS. Les solutions ou bibliothèques non compatibles de Linux, ou de Windows par exemple ne sont pas retenues dans cette étude : en vue de conserver la caractéristique multiplateforme du moteur.

Bien que la focale soit réglée sur les protocoles ouverts et standardisés, encore faut-il que les bibliothèques logicielles existent et en langage C ou C++ (langage du moteur Unreal Engine), selon un code multiplateforme. Il serait trop long et couteux d'avoir à développer de telles librairies.

Les bibliothèques qui satisfassent les critères de faisabilité et de compatibilité peuvent être classées selon leur facilité d'intégration, évaluable en temps d'implémentation et effort de conception.

### 1.1.3. De la scalabilité

Unreal Engine n'est pas conçu pour des applications critiques dont les exigences de sûreté de fonctionnement sont élevées. Cependant, cela ne doit pas empêcher de pouvoir garantir un certain niveau de service (SLA – Service-Level Agreement). Le SLA garantit au titulaire du contrat un niveau de qualité selon différents critères dont la disponibilité des serveurs, la fiabilité, le temps de réponse, etc. Afin de satisfaire de telles exigences, la scalabilité constitue une réponse largement employée. Il s'agit d'une caractéristique qui prend sens au niveau des serveurs et non des postes clients employés par les utilisateurs.

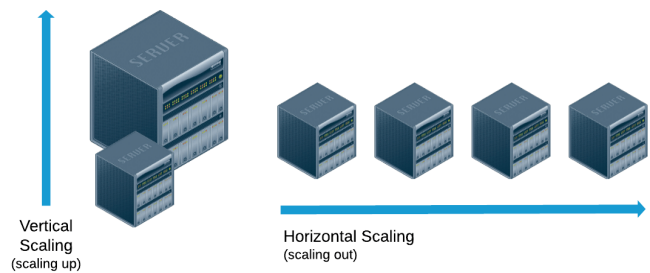

*Figure 1 : illustration de la scalabilité (visuel www.section.io)*

> La scalabilité verticale consiste à augmenter la puissance des serveurs pour répondre au besoin de ressources alors que la demande augmente (généralement due à un nombre croissant d'utilisateur). Il s'agit d'agir sur le matériel. Par conséquent, la scalabilité est hors du scope de ce document.
>
> La scalabilité horizontale est basée sur l'ajout de serveurs en parallèle en vue de répartir la charge sur plusieurs machines (généralement, il s'agit de machines virtuelles ou de conteneurs) : les serveurs sont extensibles.
>
> Il est possible d'approfondir davantage ces deux concepts mais ces définitions sont suffisantes pour cette étude.

**1.1.4. De la flexibilité**

La flexibilité des solutions traduit leur capacité à s'adapter à plusieurs besoins ou formats et caractéristiques de données. Aucune solution n'est parfaite et aucune n'est adaptée à tous les besoins : tout est affaire de compromis ou de dilemme, entre la généricité d'une solution, l'effort nécessaire pour la faire correspondre à un besoin et l'effort qui sera nécessaire pour faire évoluer ce besoin, à l'avenir. Les architectures les plus flexibles et évolutives, sont aussi les plus lourdes, et chers à implémenter, à maintenir et à exploiter. Par opposition, une solution plus rigide peut remplir le besoin à un instant mais nécessiter un redéveloppement significatif pour satisfaire certaines évolutions.

## 2. LES ARCHITECTURES

### 2.1. Le modèle client-server d'Unreal Engine

Dans l'univers des jeux vidéo, le multijoueur est devenu la norme depuis déjà plus d'une décennie. Le développement d'un jeu multijoueur implique de créer des fonctionnalités réseaux plus ou moins avancées, fonction des besoins, du type de jeu et évidemment, du nombre de joueurs en simultanés.

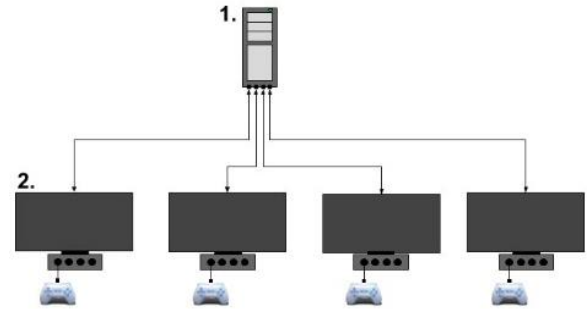

Lors d'une partie multijoueur, les données sont transmises entre tous les joueurs (les machines), via la connexion réseau (souvent Internet). Ces données renseignent sur l'état du jeu afin que toutes les machines sur le réseau puissent reconstruire un environnement cohérent face à la scène qui est en train de se dérouler. Il s'agit ainsi de communiquer, par exemple, les déplacements des personnages, les actions qu'effectuent ces personnages (comme ramasser un objet) et tous les évènements susceptibles de faire évoluer l'état du jeu.

Le présent titre vise à analyser les modes multijoueur d'Unreal Engine en vue de les exploiter pour interconnecter des systèmes tiers ou injecter des données.

Unreal Engine offre 3 possibilités pour concevoir une application multijoueur (qu'il s'agisse d'un jeu, d'une simulation 3D ou tout autre utilisation finale) : le mode local (Standalone), le mode serveur d'écoute (Listen Server) et le mode serveur dédié (Dedicated Server). Ces 2 derniers modes utilisent le modèle client-serveur. Nous nous intéressons au mode serveur dédié.

Dans le mode serveur dédié, Unreal Server, héberge la partie et assure toutes les fonctions propres au serveur : réplication des évènements sur le réseau. Le serveur dédié ne gère aucun rendu graphique. Les clients se connectent tous au serveur pour participer à la même session. Ce mode paraît plus intelligible et les rôles de chacun sont bien répartis. Ce mode est le plus adapté pour rendre une application 3D multi-utilisateur.

### 2.2. Les technologies sous-jacentes du modèle client-serveur

Instruit des différents cas d'utilisation développés par Epic Games, il est nécessaire de comprendre les techniques sous-jacentes et d'évaluer notre capacité à les exploiter.

La documentation officielle d'Unreal Engine détaille très bien les mécanismes qui suivent, en voici une synthèse volontairement simplifiée et suffisante pour la suite. Il s'agit des mécanismes de réplication et d'appels de procédure distante (RPC – Remote Procedure Call).

#### 2.2.1. La réplication

Le mécanisme de réplication reproduit les informations sur l'état du jeu vers les différentes machines. Unreal Engine propose plusieurs fonctions de réplication mais le concept reste le même : il s'agit de transmettre l'état d'un objet, d'un composant ou d'une variable aux autres utilisateurs sur le réseau. Nous pouvons par exemple choisir de répliquer le personnage d'un joueur afin que tous les utilisateurs voient les mouvements de ce dernier.

L'utilisation de la réplication est très simple pour le développeur. Un simple menu déroulant permet de répliquer presque n'importe quels éléments dans Unreal Engine. Une petite action qui doit entrainer de grandes préoccupations : il est important de déterminer la pertinence de répliquer un élément sous peine de subir un coût réseau significatif.

#### 2.2.2. Remote Procedure Call

Les appels de procédures distantes permettent d'appeler une fonction sur un ordinateur distant de la même manière que si la fonction était disponible en local. En synthèse, RPC permet d'effectuer ce que la réplication ne fait pas : c'est-à-dire exécuter une procédure spécifique, sur une machine spécifique. Là où la réplication est un mécanisme global, RPC permet de gérer finement des évènements distants.

> Le concept de Remote Procedure Call n'est pas spécifique à Unreal Engine et a été spécifié en 1988 [6].

Le point-clé ici est qu'il n'existe aucune implémentation standard uniforme pour les appels RPC. Concrètement, les bibliothèques qui implémentent du RPC ne sont pas nécessairement compatibles entre elles.

Les mécanismes de réplication et RPC sont donc des mécanismes à considérer internes à Unreal Engine : des mécanismes pertinents pour transférer des données entre un serveur UE et un client UE. Le code source d'Unreal Engine étant accessible (Open Source), il est possible de le réutiliser pour implémenter RPC dans un autre système tiers, et rendre les échanges possibles entre ce système tiers et Unreal Engine : oublions rapidement cette hypothèse absurde par l'énergie et le temps qu'elle nécessiterait d'engager. La solution pour intégrer des données externes ne se trouve donc pas dans les mécanismes multijoueur intégrés par Epic Games.

> Synchronisation du temps : dans le cas de simulations temps réel où des événements transitent à travers le réseau, il est nécessaire de disposer d'outils pour synchroniser les horloges des différentes machines : afin que ces dernières partagent la même référence. L'objectif d'une telle synchronisation est d'éliminer le décalage de temps, dû au réseau.
>
> Unreal Engine implémente un système de synchronisation de temps qu'il est nécessaire d'utiliser pour réduire l'impact des décalages (lags) réseau.

## 2.3. La scalabilité

Dans le cadre d'application 3D ou de simulateurs largement déployés, où il est susceptible d'accueillir un nombre croissant de données et d'utilisateurs, il devient primordial que le produit réalisé via Unreal Engine, pour sa partie serveur, soit extensible horizontalement (un rappel de la définition est disponible au titre 1.1.3). Tuons le sujet dans l'œuf : Unreal Engine n'est pas « scalable » horizontalement.

Le modèle client-serveur d'Unreal Engine est le modèle de prédilection pour rendre une application multi-utilisateur, en particulier via le mode serveur dédié. Pour fonctionner, ce mode utilise les systèmes de réplication et RPC. Ces systèmes se basent sur le protocole UDP et sont stateful (avec état) : en particulier la réplication, qui conserve l'état de l'application au cours du temps.

> Pourquoi Unreal Server est stateful ? Imaginons une session à 3 utilisateurs, matérialisés par des avatars. Nous sommes dans le cas d'un serveur dédié et ces 3 utilisateurs y sont connectés. Par défaut, partons du principe que tous les avatars ont l'apparence du très connu Mario . L'un des personnages, se transforme en Luigi. L'application de cet utilisateur envoie alors au serveur l'information « transformé en Luigi ». Le serveur retransmet l'information aux 2 autres utilisateurs qui voient désormais leur camarade sous l'apparence de Luigi . Si le serveur ne conservait pas l'état du système, et qu'un 4ème utilisateur rejoint la session, ce dernier ne saura pas qu'un des protagonistes s'est transformé en Luigi : il ne verra que des Mario . En réalité, Unreal Server conserve la connaissance de l'état du système afin de la communiquer aux nouveaux arrivants.
>
> Cet exemple est volontairement simplifié. En effet, un serveur Unreal pourrait être stateless s'il récupérait, par exemple, l'apparence des personnages depuis une base de données, lorsque nécessaire. Cependant, les temps de traitement ne sont alors plus les mêmes et dans ce cas, le service ne serait pas en mesure d'assurer la meilleure latence possible.

La condition essentielle pour qu'un serveur soit évolutif horizontalement est qu'il soit stateless (sans conservation d'un état entre 2 requêtes de clients). Une condition souhaitable serait également qu'il utilise HTTP (basé sur TCP, à la place d'UDP). Ces 2 contraintes s'expliquent en vue de satisfaire les exigences de compatibilité avec les répartiteurs de charge (Load Balancer).

Les Load Balancer servent à répartir les requêtes des clients entre les différents serveurs. Il existe plusieurs algorithmes pour répartir la charge, les plus simples étant le tourniquet (Round-robin) qui effectue une rotation circulaire. Pour fonctionner, tous les serveurs doivent être identiques : ils ne peuvent pas se situer dans un état différent.

Concernant le protocole UDP, ce dernier ne fonctionne pas en mode connecté et ne propose aucun mécanisme de contrôle ou d'acquittement. Bien qu'il existe des Load Balancer qui fonctionnent en UDP, ils nécessitent des architectures complexes pour fonctionner. UDP n'embarque pas suffisamment d'information pour que le Load Balancer puisse, avec certitude, retransmettre les trames au serveur adéquat. Sur le réseau, vu du Load Balancer, toutes les trames sont indépendantes. Pour identifier une connexion le Load Balancer peut uniquement se baser sur le couple adresse IP / Port contenu dans les trames. Un message qui arrive d'un client et qui a été fragmenté en plusieurs trames, devrait ainsi pouvoir être routé par le Load Balancer vers le même serveur qui reconstituera le message. Le couple adresse IP / Port de la première trame sert d'identification pour les suivantes. Avec cette méthode, il n'est pas possible de distinguer les utilisateurs qui utilisent la même adresse IP (ce qui est généralement le cas lorsqu'ils se trouvent sur le même sous réseau, en version 4 du protocole IP). De plus, de nos jours la majorité des fournisseurs d'accès Internet affectent une adresse IP dynamique à leurs clients : adresse IP qui change régulièrement en fonction d'un certain nombre de critères. Ces contraintes pèsent en termes d'architecture et de configuration, et d'autres inconvénients sont à prendre en compte lors du choix d'un Load Balancer sur UDP. Bien qu'il existe des cas d'usages correspondant bien à l'utilisation de ces Load Balancer, ces derniers ne sont pas idéaux pour les applications 3D.

Le serveur dédié d'Unreal Engine, parce qu'il est stateful et qu'il emploie le protocole UDP pour le transport des données entre le serveur et les clients, n'est donc pas « scalable » en l'état.

Les capacités d'extension du moteur, notamment via le langage C++, permettront d'étendre les possibilités du moteur.

## 2.4. L'impact du contexte de réalité mixte sur l'architecture

### 2.4.1. L'analyse de la réalité mixte

L'utilisation de la réalité mixte, et en particulier des casques de réalité virtuelle apportent une dimension supplémentaire à l'expérience vécue.

La réalité virtuelle apporte plusieurs avantages :

Une meilleure prise en compte de l'environnement et une ergonomie augmentée ainsi qu'une diminution de l'effet tunnel de l'utilisateur (par opposition à l'usage d'un écran). Les gestes sont plus naturels, il est par exemple plus évident de tourner la tête que d'utiliser des touches pour orienter la vue comme c'est le cas sur un écran. Certains contrôles deviennent usuels et ainsi l'ergonomie est améliorée, tout comme le temps d'apprentissage de l'application elle-même par l'utilisateur. De surcroit, l'utilisateur est déchargé de l'utilisation de touches pour orienter la vue ce qui réduit sa charge mentale. Ces effets produisent des opportunités dans le cadre de la formation ou du télé-pilotage en temps réel.

Dans le cadre du pilotage d'aéronef, notamment en règles de vol à vue (VFR – Visual Flight Rules) ou en pilotage manuel, la visualisation des trajectoires est plus précise et fidèle pour l'utilisateur, que lors de l'usage d'un écran. La vision stéréoscopique contribue à une meilleure appréciation des distances et des trajectoires. Ces avantages peuvent également être pertinents dans le cadre d'usage de drones terrestres.

Une formation est d'autant plus efficace si elle est réaliste. La réalité virtuelle améliore l'immersion et la sensation de réalisme. Cette dernière constitue donc une opportunité d'améliorer la formation dans plusieurs domaines, à moindre coût.

Un simulateur accessible en réalité virtuelle améliore la fidélité du jumeau numérique offerte par le simulateur, en particulier pour l'œil humain.

En revanche, comme toute technologie elle est également porteuse d'inconvénients :

L'utilisateur est plus en proie à la fatigue, précisément, lors de séances prolongées. Cette dernière peut être d'ordre oculaire et provoquer des maux de tête, mais aussi liée au poids du casque VR (Virtual Reality) utilisé. Des douleurs aux cervicales peuvent notamment apparaître.

La majorité des utilisateurs non habitués ni entrainés à la VR peuvent ressentir des sensations de vertiges proches du mal des transports. Il s'agit essentiellement de cinétose (bénigne) provoquée par une incohérence entre les informations d'orientation et d'inclinaison captées par l'oreille interne et la vision de l'utilisateur. Cette sensation disparaît avec l'entrainement de l'utilisateur.

Les casques VR ne sont pas toujours adaptés et compatibles avec les personnes amétropes ou souffrantes de handicap visuel.

### 2.4.2. Les impacts sur l'architecture

Maintenir une application en réalité virtuelle agréable implique de maintenir une fréquence de rafraichissement d'images suffisantes. Cette fréquence d'image doit permettre d'apporter une expérience utilisateur agréable mais aussi et surtout d'empêcher les sensations de vertiges. Il est communément admis que la fréquence de référence dans l'industrie de la réalité virtuelle est de 90 IPS. En dessous de cette valeur, l'immersion peut être affectée. Concernant les vertiges, difficile d'en établir un seuil tant les individus sont différents. Estimant être un individu plutôt tolérant, les sensations de vertiges apparaissent dans mon cas sous 30 IPS. Outre la fréquence d'image, la latence joue aussi sur le confort et l'apparition des vertiges. Par exemple, l'utilisateur ne doit pas percevoir de latence entre l'instant où il tourne la tête et l'instant où le système génère les images cohérentes avec la nouvelle orientation.

Les contraintes de latence et de maintien d'une fréquence d'image à 90 IPS pèsent essentiellement sur la programmation de l'application 3D, côté client et sur la puissance de la machine. Cependant, il faut garder cette exigence en vue. En effet, l'élément affiché du côté de l'utilisateur provient souvent de l'agrégation d'une ou plusieurs données, transmises par le serveur. Dans la majorité des cas, l'application de l'utilisateur se contente d'afficher des modèles 3D et d'en actualiser les modèles (positions, états). Un rafraichissement périodique d'un modèle 3D dans la scène, à une faible fréquence, n'entrainera pas plus de cinétose qu'habituellement. De surcroit, il est souvent possible d'extrapoler la position des objets en vue d'offrir une expérience fluide. Cependant il faut prêter attention aux environnements sombres et lorsque les modèles 3D affichés prennent une place importante dans le champ de vision de l'utilisateur. Il s'agit d'un cas où la fluidité des mouvements doit être préservée. Un maximum doit être entrepris pour que le champ de vision de l'utilisateur ne paraisse pas figé, quelles que soient les circonstances. Cette contrainte peut engendrer un impact sur l'implémentation, côté client, mais aussi côté serveur.

## 2.5. Création d'une extension gRPC

La liste des API (Application Programming Interface) et extensions déjà développée est riche, notamment via l'extension VaRest qui offre un accès aux API REST aujourd'hui largement déployés. Cependant, il manque au catalogue un candidat intéressant et plébiscité : gRPC.

> Là où REST favorise la simplicité, gRPC plaide pour la performance. Les API REST font transiter des messages textuels, sous les formats JSON ou XML généralement. Ces messages sont donc naturellement plus volumineux et nécessitent un traitement par les machines. La charge utile est plus faible qu'avec gRPC mais les API REST sont très flexibles.
>
> Par opposition, gRPC est stricte dans sa spécification des interfaces. Il tend à minimiser la quantité de données transmises pour en accroitre la rapidité. La liste des langages compatibles est large (11 langages à ce jour) et sa mise en œuvre légèrement moins triviales qu'un API REST. gRPC s'appuie sur HTTP version 2 et il n'est pas systématiquement compatible avec les navigateurs [7]. Le code de gRPC est accessible et gratuit. Il embarque une licence Apache qui permet une utilisation privée, commerciale et sans contamination. gRPC est une bibliothèque et sa spécification est ouverte. Il s'agit d'un projet de la CNCF (Cloud Native Computing Fundation) et initié par Google.
>
> gRPC n'est donc pas concurrent des API REST mais répond à un besoin différent.
>
> Le plus sérieux concurrent de gRPC est SignalR. Les performances de ces 2 bibliothèques sont comparables et les différences sont plus subtiles. Personnellement, je pense gRPC intéressant car il propose un standard plus ouvert. SignalR implique d'utiliser l'environnement .NET Core de Microsoft ce qui limite les langages compatibles. Note importante : gRPC utilise HTTP version 2 et SignalR est compatible de HTTP version 1. Il ne faut donc pas se jeter à corps perdu sur gRPC, sans analyse, dans le cas où les clients utiliseront un navigateur.

Il n'existe pas d'extension clé en main pour intégrer gRPC à Unreal Engine. Les articles qui proposent son intégration à Unreal Engine ne sont pas très nombreux et il y a de nombreuses complexités à franchir pour intégrer correctement une librairie externe C++ dans Unreal Engine. La version 5 étant disponible récemment (à la date des travaux), aucun article aborde cette version. J'ai réalisé pour l'occasion une extension fonctionnelle permettant d'intégrer gRPC dans Unreal Engine. Plusieurs travaux m'ont permis d'aboutir :

• L'article « UE4.26 with gRPC » [8] propose un processus à suivre. J'ai rencontré beaucoup de difficulté à aboutir via cette méthode, en raison des directives de pré-compilation (#pragma) à insérer dans le code et qui ne sont pas toujours pérennes face à l'évolution des versions (évolution des versions de gRPC et d'UE). De surcroit, la méthode implique de modifier des fichiers qui sont générés par gRPC (Protocol Buffer exactement). Idéalement, il vaut mieux s'abstenir de modifier les produits de l'étape de génération de Protocol Buffer. Cet article, écrit en japonais (et j'en remercie l'efficacité des services de traduction en ligne), constitue cependant une excellente base de travail pour l'intégration de gRPC.

> Protocol Buffer est le format de sérialisation des données utilisés, par défaut, dans gRPC.

• Le code source « InfraworldRuntime » propose une extension gRPC pour UE. Cependant, la dernière mise à jour date du 16 juin 2020, gRPC et UE ont largement évolué depuis.

• L'article « Microservice » [9] propose d'embarquer un serveur UE dans un conteneur microservice. L'opération comprend une intégration de gRPC via le gestionnaire de paquets Conan et l'emploi du langage Python. Ces recherches extrêmement intéressantes apportent un processus et des outils complets pour compiler des librairies externes et les intégrer à Unreal Engine. Ce projet va bien au-delà de notre besoin mais prouve, via ce qu'il annonce, que les limites peuvent être franchies. Il nécessite une bonne appropriation du gestionnaire Conan et de concepts clés sur le langage C++ et Unreal Engine.

Mon approche ici a été plus directe et vise à proposer une extension gRPC compatible des standards Unreal Engine et s'intégrant comme toutes les extensions du moteur.

**2.5.1. Procédé de création de l'extension**

La première étape consiste à compiler les sources de gRPC. Il n'y a ici pas de difficulté particulière hormis celle de maîtriser les opérations inhérentes au langage C++. La documentation officielle de gRPC explique la démarche, pour plusieurs environnements et via plusieurs outils. Cette étape produit les librairies statiques (« .lib » sous Windows et « .a » si l'on utilise Linux) qu'il faut bien identifier.

Epic Games fournit des éléments de documentation sur l'intégration de librairies extérieures, ainsi qu'un canevas d'extension bien venue (Third-Party Libraries). Cette référence apporte une vision globale sur la manière d'incorporer une librairie tierce mais sans aborder les problèmes réels qui sont susceptibles de se produire.

La méthode mise en avant par le canevas consiste à créer un sous-projet au projet de Plugin (= extension), aboutissant à une compilation en 2 étapes : 1/ la librairie extérieure (ThirdParty). 2/ le projet UE.

**Librairie ThirdParty**

La librairie du dossier ThirdParty contiendra le code natif C++/gRPC. C'est dans cette librairie qu'il faut écrire le code C++ qui fournit toutes les fonctions gRPC utiles au Plugin et propre au projet développé. Elle est nommée ThirdParty dans la mesure où elle est extérieure à l'application Unreal Engine. Un fichier identifié servira d'interface et de point d'entrée pour l'extension.

Le canevas encourage à effectuer une compilation dynamique en vue de générer une librairie « .dll » (« .so » sous Linux) : il faut s'abstenir et préférer une compilation statique, en particulier sous un environnement Windows.

En effet, gRPC ne recommande pas la compilation via DLL (Dynamic Link Library) car il existe certains inconvénients connus concernant le fonctionnement des DLL C++ sous Windows. Par exemple, il n'y a pas d'ABI (Application Binary Interface) C++ stable et il n'est pas possible d'allouer de la mémoire en toute sécurité dans une DLL et de la libérer dans une autre. Passer à côté de cet impératif entraîne des crashs imprévisibles dont l'origine est impossible à identifier via le débogage.

De surcroit, la compilation via DLL risque d'entrainer des problèmes lors de l'utilisation de types C++ dans le fichier d'interface, comme par exemple l'utilisation de types std::string de la STL (Standard Template Library). Cette librairie, standard dans son API, ne possède pas d'ABI standard. En effet, si la librairie ThirdParty est liée (étape d'édition de liens après la compilation) avec une bibliothèque STL C++ dont l'implémentation est différente de celle utilisée lors de l'édition de liens du Plugin, alors cette dernière étape échouera sur une erreur de type : symbole non défini. Concrètement, Unreal Engine utilise par exemple, la librairie « libc++ » sous Linux et non « libstdc++ » que l'on trouve également installée sur ces systèmes. Par conséquent, une édition de liens de la librairie ThirdParty (le code natif gRPC qu'il faut écrire) avec « libstdc++ » entrainera une erreur à la compilation de l'extension. Sous Windows, Unreal Engine utilise MSVC (MicroSoft Visual C++).

Les fichiers de configuration d'UnrealBuildTool doivent ainsi être édités en phase avec cette stratégie. La compilation par défaut est : gRPC statique -> thirdparty dynamique > dynamique. Il faut la modifier comme suite : gRPC statique -> thirdparty statique -> plugin statique.

**Librairie du Plugin**

L'étape de compilation de l'extension est facilitée par le respect des étapes précédentes. Une mauvaise configuration peut engendrer de très nombreux « Warnings » ou erreurs à la compilation indiquant des problèmes d'interposition de symboles. Epic Games fournit les macros « THIRD_PARTY_INCLUDES_START » et « THIRD_PARTY_INCLUDES_END » qui permettent d'encapsuler un code externe et éviter les conflits de redéfinition. De plus, Unreal Engine traite de nombreux Warnings comme des erreurs alors que ce n'est pas toujours le cas des librairies extérieures. Cependant, ces macros ne produisent pas de miracles et une mauvaise chaîne de compilation peut mener à un problème insolvable. Cette méthode permet également d'éviter d'utiliser les macros ci-dessus dans le code généré par Protocol Buffer : ce qui ne constitue pas une bonne pratique et peut mener à un processus de développement hasardeux.

Il est à noter que cette procédure fonctionne car l'ensemble de toutes les librairies est recompilé à partir des sources, en partant des sources de gRPC. Dans le cas d'intégration d'une bibliothèque déjà compilée et dont on ne dispose pas du code source, la solution peut s'avérer plus compliquée à trouver.

L'étape de compilation franchie, il est enfin possible de se lancer sur le développement d'une architecture logicielle saine et sérieuse. La conception de ce Plugin consiste premièrement à implémenter un « Wrapper » ; soit le commun mais efficace patron de conception « Adapteur ». Ce dernier est largement documenté sur Internet et pour de nombreux langages. Les programmeurs qui souhaiteront se lancer dans un tel développement ne feront pas l'impasse d'associer à leur architecture, le patron de conception « Observateur » pour faire remonter des messages depuis gRPC vers le monde 3D. Ce modèle d'implémentation permet de minimiser le couplage entre les composants logiciels.

### 2.5.2. Les architectures possibles avec gRPC

gRPC emploie un mécanisme entièrement stateless. Il utilise Protocol Buffers comme langage de description d'interface (IDL – Interface Definition Language).

Il permet d'échanger des structures de données avancées entre les stubs (les clients) et le serveur.

gRPC se base sur un modèle client-serveur. Le serveur peut être embarqué dans un composant indépendant, via un des 11 langages compatibles comme Python. Ceci permet de bâtir une architecture scalable qui s'appuie sur un backend (Figure 2).

L'extension développée propose les 2 options : soit d'intégrer le client gRPC dans une application Unreal Engine, soit d'y intégrer le serveur gRPC. Cette deuxième option autorise la création d'architectures plus directes telles que sur la Figure 3 où un client gRPC (par exemple Matlab Symulink) se connecte au serveur embarqué dans une application 3D. Finalement toutes les options sont réalisables comme intégrer un client gRPC dans Unreal Engine et se connecter aussi en direct à une autre application, qui embarquerait le serveur gRPC.

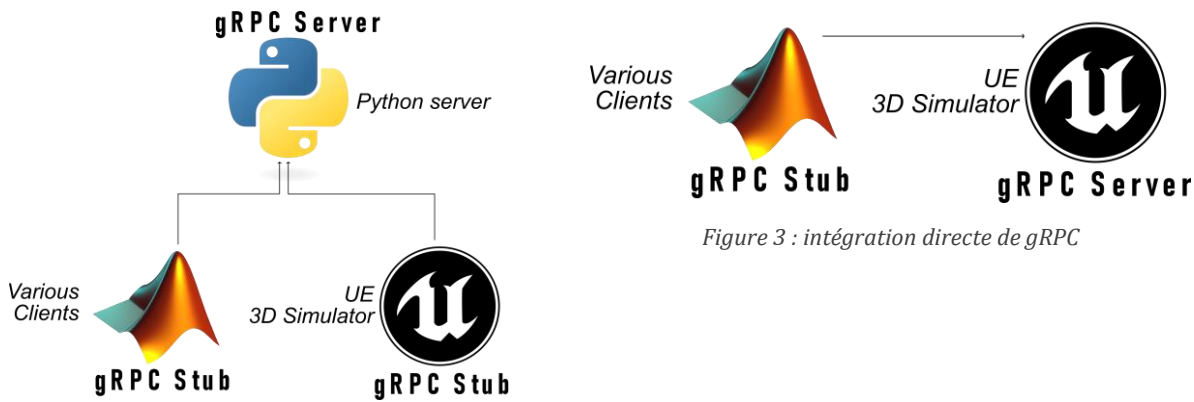

*Figure 3 : intégration directe de gRPC*

*Figure 2 : intégration de gRPC via un backend*

Une connexion gRPC est toujours initiée par le client. Cependant, là où les connexions en API REST sont unidirectionnelles, elles peuvent être bidirectionnelles avec gRPC. Un client qui se connecte à un serveur peut laisser ouvert un « stream » en vue de recevoir des évènements, continues ou non, qui proviennent du serveur. Cette mécanique peut paraître disgracieuse à implémenter car user de boucle mais la fonctionnalité est là.

L'interface entre les composants logiciels se défini dans un fichier « .proto » (Protocol Buffers) agnostique du langage informatique utilisé. Le fichier « .proto » constitue le contrat entre les différentes applications. Une fois le fichier proto établi et la compilation fonctionnelle, les étapes restantes relèvent du génie logiciel pur et nécessitent une connaissance du C++/Blueprint dans Unreal Engine. Il s'agit d'implémenter 2 « wrapper » C++. Le premier encapsule gRPC dans un fichier interface, exposé par la librairie ThirdParty à développer. Le 2ème « wrapper » encapsule ce fichier d'en-tête dans une classe Actor propre à Unreal Engine. Ici, la programmation concurrente (multithreading) est nécessaire pour ne pas bloquer le rendu du moteur. Unreal Engine fournit la classe FRunnable pour créer un fil d'exécution séparé du fil d'exécution principale.

La lecture-écriture de données, entre le Thread et un Actor (composants généralement affichés dans une scène 3D), peut s'effectuer via l'implémentation du patron de conception « Observateur », entre les composants gRPC et le Thread. La remontée des données dans un Actor, doit employer le conteneur TQueue<T> conçu pour les accès concurrents et complémentaire du système Delegate d'UE. Ce dernier permet de faire remonter un événement au fil d'exécution principal. Attention à découpler ce Thread du fil d'exécution principal grâce aux AsyncTask d'UE, lors de l'émission d'un l'événement.

### 2.5.3. Mise en œuvre dans Unreal Engine

Le Plugin une fois établi permet d'être utilisé en langage Blueprint dans Unreal Engine.

La Figure 4 illustre un moyen de démarrer un client gRPC dans Unreal Engine, qui ouvre un flux « stream » entrant.

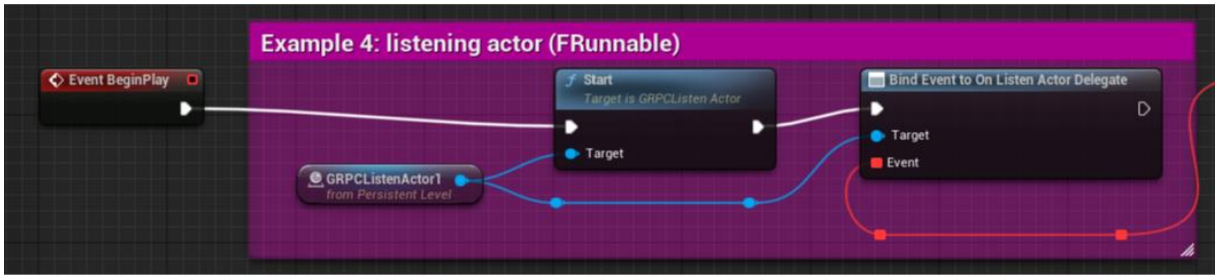

*Figure 4 : démarrage du serveur gRPC, Blueprint*

La Figure 5 démontre un moyen pour réceptionner les messages.

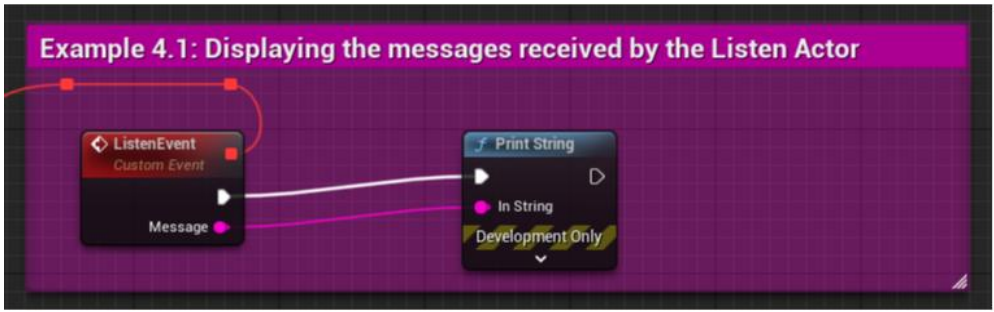

*Figure 5 : réception des messages gRPC, Blueprint*

Via un tel Plugin, l'utilisation de gRPC semble être transparente pour le développeur du simulateur ou de l'application 3D. Il est à noter que cette extension doit être vue comme un canevas plutôt qu'un Plugin clé en main. En effet, par la nature des bibliothèques gRPC/Protocole Buffers, Il reste nécessaire d'adapter le code source du Plugin afin que ce dernier corresponde à ses besoins.

Il s'agit là d'une bonne alternative aux API REST, qui permet de gagner en performances tout en conservant la caractéristique de scalabilité.

# 3. RECOMMANDATIONS

## 3.1. Simulateur Cloud Native

Le présent titre propose une architecture plus générique, pouvant répondre à une interconnexion d'un simulateur avec un SI ou tout autre composant (Hardware In the Loop, Plateforme IoT « Internet of Things », …).

Une telle architecture présente les avantages d'améliorer la disponibilité du service (le simulateur), de proposer de la redondance, d'encaisser la charge face à une augmentation du nombre d'utilisateurs ou de la quantité de données à intégrer dans le système. La performance de l'ensemble est alors grandement améliorée.

Un simulateur immersif et augmenté, accessible à plusieurs services d'une entreprise ainsi qu'à ses clients et répondant à un SLA (Service-Level Agreement) nécessite de déployer une architecture complexe.

Les exigences les plus extrêmes de faible latence ne pourront pas être rendues conformes ici : une architecture Cloud engendre un temps réseau, lié à la distance entre l'utilisateur et le data center et un temps de traitement, lié aux différents serveurs à traverser. C'est peut-être la seule caractéristique technique difficile à surpasser, par opposition à une connexion directe.

> Cette architecture ne traite pas certains points, qui sont : le site de reprise après désastre (Disaster Recovery site), les composants de sécurité pure tel que le SSO (Single Sign-On), l'Active Directory (services Lightweight Directory Access Protocol) ou encore les systèmes de détection d'intrusion qui sont généralement en dehors du périmètre applicatif.
>
> Le coût des différentes solutions n'est pas abordé dans ce document. Il reste cependant utile de garder à l'esprit, qu'outre le fait que le nombre brut de machines virtuelles augmente le coût, c'est aussi le cas du nombre de solutions / logiciels à déployer, du nombre de technologies différentes à maîtriser (langages informatiques, clusters, redondance active-active vs. active-passive…). Tous ces éléments influencent le coût d'achat et le coût d'exploitation.

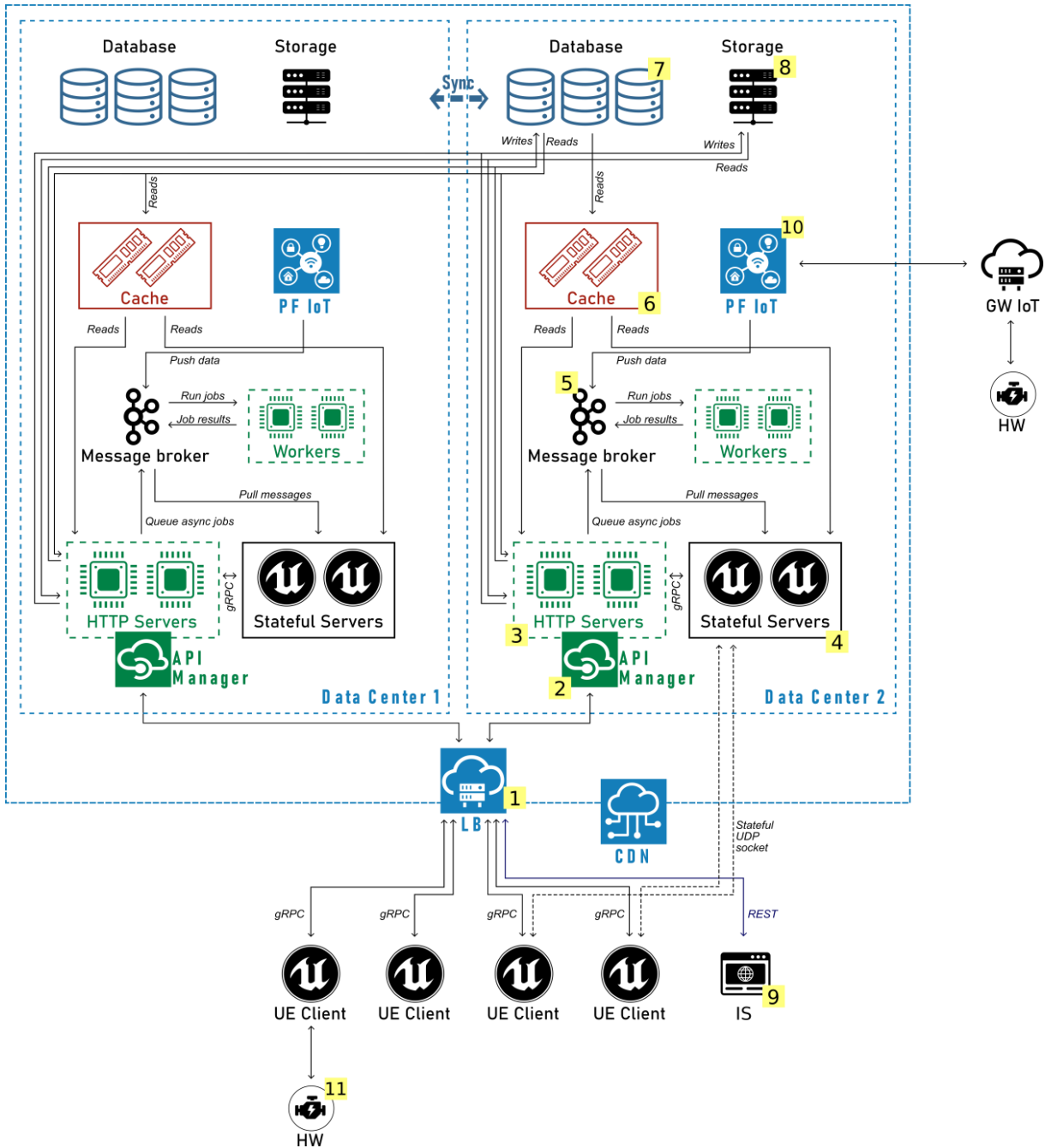

*Figure 6 : architecture Unreal Engine Cloud Native*

Cette architecture, créée à partir des différentes analyses de cette étude, se veut volontairement riche. Elle s'inspire et modifie une introduction publiée par [10]. Elle peut être vue comme une offre à tiroirs dont on peut retirer ou remplacer les composants. Elle propose une haute disponibilité (actif/actif) avec duplication complète sur 2 centres de données distincts.

Cette architecture évolutive horizontalement, peut prendre en charge un large spectre de fonctionnalités tel que l'authentification des utilisateurs, le chat, le multi-utilisateur temps réel, le contenu généré par l'utilisateur, le téléchargement des exécutables et leur mise à jour, la création de tableaux de bord, l'intégration de données issues d'objets connectés ou d'un système d'information tiers, … En fonction des besoins, certains composants peuvent ainsi être économisés. Elle ne colle pas précisément à un cas d'usage mais peut servir de base de travail pour une application 3D industrielle. Non forcément destinée qu'aux cas où le nombre d'utilisateurs est grandissant, elle est aussi en mesure de convenir aux caractéristiques du Big Data (gros volumes de données, hétérogénéité et haute vitesse) et combler un besoin d'interopérabilité à grande échelle. Il s'agit ainsi d'adresser des thématiques comme l'intégration de données massives dans une application 3D, un simulateur ou proposer une application en haut disponibilité.

| N°  | Description |
| --- | --- |
| 1   | La plateforme du simulateur est répartie sur 2 centres de données distincts. La distance admissible entre les 2 centres de données dépend de la technologie réseau employée et du type de base de données déployée. Le Load Balancer est ici compatible HTTP/2 car il doit permettre de prendre en charge gRPC. L'utilisation d'un CDN (Content Delivery Network) est recommandée mais facultative. Il permet d'améliorer le chargement de certaines ressources statiques pour les utilisateurs : ce qui prend sens si ces derniers sont éloignés du Data Center, ou irrégulièrement répartis. |
| 2   | L'API Manager est facultatif. Il agit comme un proxy entre le client et le service web et apporte des fonctionnalités de sécurité (authentification) ou de gestion de niveaux de qualité de service. Remarque : il n'agit pas sur les protocoles bâtis sur UDP. Il existe plusieurs solutions paramétrables comme WSO2 (logiciel Open Source), Tyk, Axway ou encore AWS Gateway. |
| 3   | Les serveurs HTTP sont stateless et scalables dynamiquement. Ils implémentent des web services gRPC ou REST. Ces serveurs doivent servir à traiter un maximum de fonctionnalités, tel que la connexion, la messagerie instantanée, les statistiques, la collecte de données et métriques, les événements asynchrones et plus généralement tous les événements qui ne répondent pas à des contraintes temps réel. Un simulateur ou une application 3D industrielles qui a peu ou pas de contraintes de latence pourrait donc se passer des serveurs stateful UE et utiliser exclusivement des serveurs HTTP. Ceci simplifie énormément l'architecture, réduit son coût et améliore sa scalabilité. Les serveurs HTTP peuvent être codés dans de nombreux langages comme Python, C#, Javascript, Go, Java, C++, … |

> La synchronisation des horloges n'est pas présentée dans cette architecture. Si les serveurs Unreal Engine sont utilisés, il reste possible d'utiliser l'implémentation intégrée au moteur. Dans le cas contraire, il faut garder à l'esprit qu'une méthode de synchronisation de temps peut s'avérer nécessaire, suivant le cas d'usage.

| | |
| --- | --- |
| 4   | Les serveurs stateful sont des serveurs dédiés Unreal Engine utilisant les systèmes de réplication/RPC, basés sur UDP. Ils permettent de traiter les requêtes affectées de contraintes temps réel (faible latence). Il s'agit généralement d'actions multi-utilisateurs, comme la transmission des déplacements des véhicules, des personnes ou des actions à transmettre en un minimum de temps. Ces serveurs sont stateful car ils doivent stocker l'état de l'application 3D et restituer cet état, chaque fois que possible, et rapidement. La meilleure latence est ainsi obtenue avec ces serveurs. Ces serveurs ont cependant un coût élevé car ils ne sont pas évolutifs horizontalement : par conséquent, pour accepter un nombre grandissant d'utilisateurs, il faut augmenter la puissance de la machine, ce qui coûte généralement bien plus cher. De surcroit, le nombre d'utilisateurs simultanés est limité par la capacité d'un serveur, aussi puissant soit-il. |
| 5   | Le broker de message est dédié aux tâches asynchrones. Il permet d'exécuter des tâches intensives, via les Workers, et donc le traitement d'une grande quantité de données en parallèle des serveurs supportant les utilisateurs. De plus, par l'usage de connexion asynchrone, il est découplé des autres serveurs et assure donc son indépendance. La puissance de traitement du broker de message en fait un composant essentiel. Plusieurs produits existent dont les très connus Apache Kafka et RabbitMQ. Apache Kafka est un compagnon parfait pour du Big Data et peut traiter plus d'un million de messages par seconde (lorsque déployé en cluster). RabbitMQ autorise un volume de traitement plus faible mais offre la meilleure latence. RabbitMQ permet le Pull et le Push de données là où Apache Kafka ne permet que le Pull. Une synchronisation entre les clusters des différents data centers peut être mise en place. |

> **Pull**, c'est le client qui vient récupérer les données auprès du broker de message. **Push**, c'est le broker de message qui notifie le client de l'évènement.

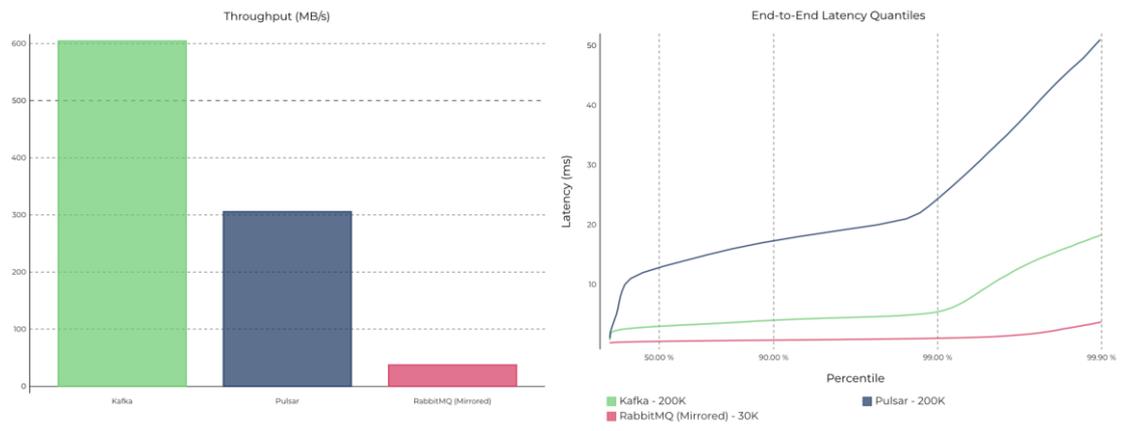

*Figure 7 : débit de données et latence, Kafka, Pulsar et RabbitMQ [11]*

| | |
|---|---|
| 6 | L'ajout d'un système de cache est facultatif et peut s'effectuer fonction de l'augmentation de la charge sur les serveurs et en particulier ceux de la base de données. Il s'agit de charger en mémoire RAM les données ou les résultats de requêtes fréquemment effectuées afin de les servir plus rapidement. Il existe plusieurs solutions comme Memcached ou Redis. Une synchronisation entre les clusters des différents data centers peut être mise en place.<br><br>> Ici, Redis ne sert pas à persister les données. Il n'agit donc pas comme une base de type CP (Consistante et Tolérante aux pannes) mais plutôt comme un système AP (Disponible et Tolérente aux pannes). |
| 7 | La base de données constitue bien souvent le goulot d'étranglement d'un système d'information. Plusieurs options sont possibles, un moteur de base de données relationnelle, tel que PostgreSQL ou MySQL, ou une base de données NoSQL distribuée, tel que MongoDB ou HBase. Le choix de la base dépend pour beaucoup du cas d'usage et du type de données à persister. Il est donc impossible d'en recommander une particulière, cependant, vu l'architecture proposée, il semble pertinent de choisir une base plutôt CP (Consistante et Tolérante aux pannes) ou plutôt CA (Consistante et Disponible). Evidemment, si la consistance des données n'est pas importante dans le contexte étudié, la dernière option du théorème de CAP reste valide. Bien qu'une base de données NoSQL distribuée ne soit pas aussi générique que leurs équivalents SQL, car la répartition des données ne peut correspondre qu'à un nombre de cas d'usage limité, elles deviennent cependant indispensables dans le cadre d'un système Big Data (au-delà de la centaine de terra octets).<br><br>Concernant le fonctionnement multi data centers, plusieurs décisions sont à prendre et dépendent de nombreux paramètres : le budget, le type de donnée, le cas d'usage, la solution de base de données choisie, …<br><br>Les moteurs bases de données relationnelles (SQL) ne sont généralement pas distribués : la solution la plus classique consiste à définir un serveur de base de données comme maître (utile pour les écritures et les lectures) et d'y ajouter des serveurs secondaires (utiles à la redondance et l'accès en lecture seule). Bien que la puissance d'un tel serveur maître soit élevée, sa capacité n'est pas infinie et le choix d'une base de données distribuée (NoSQL) peut devenir pertinent. Dans ce dernier cas, les architectures possibles dépendent de la stratégie de répartition (Sharding). |
| 8 | Il est nécessaire de disposer de serveurs de stockage pour servir les fichiers binaires aux utilisateurs : il s'agira des exécutables de l'application 3D, des mises à jour et tous les autres fichiers non stockables en base de données (images, vidéos, modèles 3D, …). 3 technologies sortent du lot pour cet usage : le stockage SAN (Storage Area Network), le NAS (Network Attached Storage) et le stockage objet (S3). Le SAN est l'option la plus performante et la plus chère. Le stockage objet est innovant et pertinent pour l'abstraction et la scalabilité qu'il offre. Le NAS présente le mérite d'être simple à mettre en œuvre. |
| 9 | L'intégration de données extérieures proposée dans cette architecture se situe à 3 niveaux : les données provenant d'un système d'information (IS – Information System), celles provenant d'un matériel et celles qui nécessitent un rendu en faible latence avec le simulateur.<br><br>Les serveurs HTTP présentent des API RESTful/JSON en vue d'intégrer des données au système ou d'en extraire. |
| 10 | Les plateformes IoT (Internet of Things) sont toutes indiquées pour collecter des données depuis un grand nombre de capteurs hétérogènes. Elles proposent des fonctions métiers et techniques (gestion, agrégation, analyse, sécurité, …). Leur plus-value se matérialise par le nombre important de connecteurs / protocoles que ces plateformes intègrent : les rendant compatibles avec beaucoup d'objets connectés. En plus des applications industrielles pour lesquelles elles sont destinées, elles peuvent apparaître comme de bons candidats pour la simulation HIL. Il est notamment possible d'y développer des modules pour prendre en charge un protocole propriétaire. |
| 11 | Cette architecture n'empêche pas l'interconnexion directe d'un matériel. Il reste imaginable de créer une simulation HIL en connexion directe avec un matériel et d'interconnecter cette simulation avec le système en vue d'y collecter ou d'y fusionner des données. |

*Tableau 1 : détails sur l'architecture Cloud Native*

Comme indiqué dans le tableau précédent, il s'agit là d'une architecture théorique générique et modulaire, capable d'adresser de nombreux cas d'usage mais qu'il convient d'adapter. En particulier pour atteindre un périmètre et un coût au plus juste.

## 4. CONCLUSION

L'interopérabilité d'Unreal Engine, ce qui sous-entend l'intégration de données du monde réel, ne peut être garantie qu'à condition de pouvoir y intégrer un maximum de protocoles hétérogènes. Cette condition essentielle mais pas suffisante, notamment pour satisfaire les contraintes du Big Data et assurer un service hautement disponible, doit être complétée par une caractéristique de scalabilité.

Le moteur Unreal Engine opte pour un paradigme de performance et de qualité visuelle extrême. Le choix du protocole UDP, dans ses basses couches, appuie cette orientation. Ce qui rend finalement service à la simulation et aux applications industrielles où une faible latence est recherchée.

La modularité du système d'extension permet une intégration de quasiment tous les protocoles et ainsi rend Unreal Engine interopérable. Une démonstration de création d'un Plugin gRPC a été exposée dans ce document, tout comme les pièges inhérents à l'intégration de librairies C++ externes. De nombreux autres Plugins sont d'ailleurs disponibles. Attention cependant à leur pertinence : il existe par exemple des connecteurs pour base de données SQL alors qu'il est généralement plus sécurisé et modulaire de déléguer cette interconnexion à des serveurs « backend » qui eux, exposeront une interface telle que gRPC ou REST/JSON, aux serveurs Unreal Engine.

D'autre part, l'étude de la scalabilité d'Unreal Engine est décisive pour ouvrir la voie à la haute disponibilité et aux architectures distribuées, capables d'ingérer des quantités massives de données. Ce document démontre la réalité de telles architectures. Ces dernières ne sont cependant pas sans compromis. En occultant le coût financier, le dilemme sera partagé entre une plateforme basée exclusivement sur les serveurs dédiés Unreal Engine, contre une plateforme HTTP développée spécifiquement. L'architecture proposée au titre 3.1 est hybride. Elle tente de maximiser les avantages des serveurs dédiés UE et des serveurs HTTP. Le cœur du débat est contraint par les exigences de latence. Plus ces dernières sont sévères, plus il faut s'orienter vers les serveurs dédiés Unreal Engine. A l'inverse, une latence moyenne autorise davantage d'évolutivité horizontale en s'appuyant sur des serveurs HTTP.

Pour des applications d'une centaine d'utilisateurs maximum visibles sur une même scène 3D, exploiter au maximum les serveurs dédiés d'Unreal Engine semble être la meilleure option. La productivité du framework fera gagner un temps précieux et les performances seront idéales. Ce cas semble adapté à un nombre important d'applications industrielles, notamment dans la simulation, tout en laissant la porte ouverte à un Tier 2 ou 3 évolutifs. Par opposition, l'utilisation exclusive de serveur stateless (HTTP) permettrait d'outrepasser ce maximum. Le débit réseau pouvant exploser, il constituera une grandeur dimensionnante.

Il peut demeurer difficile pour le lecteur de situer son exigence de latence : d'autant plus qu'une latence de bout en bout dépend également du choix des technologies matérielles, des technologies réseaux, des distances et des architectures techniques et logicielles implémentées. L'expérience apportée par différentes recherches m'autorise à souligner l'ordre de grandeur suivante, qui ne peut être retenu comme immuable. Ainsi, les applications inférieures à 5 ms choisiront une connexion directe sans architecture hébergée sur une infrastructure. Les applications entre 5 et 40 ms pourraient opter pour les serveurs dédiés Unreal Engine là où les exigences qui dépassent ces valeurs s'orienteront vers des serveurs HTTP.

Enfin, une hypothèse attrayante serait d'exploiter les serveurs dédiés Unreal Engine, exclusivement via le système RPC et sans utiliser la réplication. Cette hypothèse permettrait d'aboutir à une programmation d'un serveur sans état (stateless) et qui tire au maximum partie du serveur dédié d'Unreal Engine. Les états de l'application seraient persistés via une base de données et accessibles, à haute vitesse, grâce à un système de cache. Théoriquement, ce scénario permettrait d'étendre le nombre d'utilisateurs admissibles : car les serveurs dédiés Unreal Engine pourraient évoluer horizontalement. Cette supposition n'est recevable qu'après une étude d'impact du protocole UDP sur le répartiteur de charge, en particulier sur la stratégie de basculement. Ce scénario prometteur sur l'évolutivité horizontale d'Unreal Engine, mériterait recherches, approfondissement et démonstration.

L'implémentation de l'une des architectures proposées débouchera vers des nouveaux challenges à résoudre tant les branches telles que le Big Data et la simulation 3D sollicitent à la fois, la charge des machines et les flux réseaux. Le Sharding, c'est-à-dire la répartition, des utilisateurs et des données demeurera une préoccupation déterminante.

# 6. GLOSSAIRE

| | |
|---|---|
| **ABI** | **A**pplication **B**inary **I**nterface |
| **AR** | **A**ugmented **R**eality |
| **API** | **A**pplication **P**rogramming **I**nterface |
| **AWS** | **A**mazon **W**eb **S**ervice |
| **CDN** | **C**ontent **D**elivery **N**etwork |
| **CSV** | **C**omma-**S**eparated **V**alues |
| **CAP** | (Théorème) **C**onsistency **A**vailability **P**artition tolérance |
| **CNCF** | **C**loud **N**ative **C**omputing **F**undation |
| **FPGA** | **F**ield-**P**rogrammable **G**ate **A**rray |
| **GCP** | **G**oogle **C**loud **P**latform |
| **HDFS** | **H**adoop **D**istributed **F**ile **S**ystem |
| **HIL** | **H**ardware **I**n the **L**oop |
| **HLA** | **H**igh **L**evel **A**rchitecture |
| **HTTP** | **H**yper**T**ext **T**ransfert **P**rotocol |
| **IA** | **I**ntelligence **A**rtificielle |
| **IDL** | **I**nterface **D**efinition **L**anguage |
| **IETF** | **I**nternet **E**ngineering **T**ask **F**orce |
| **IEEE** | **I**nstitute of **E**lectrical and **E**lectronics **E**ngineers |
| **IoT** | **I**nternet **o**f **T**hings |
| **IP** | **I**nternet **P**rotocol |
| **JSON** | **J**ava**S**cript **O**bject **N**otation |
| **MR** | **M**ixed **R**eality |
| **MSVC** | **M**icro**S**oft **V**isual **C**++ |
| **NoSQL** | **N**ot **o**nly **S**tructured **Q**uery **L**anguage |
| **POC** | **P**roof **O**f **C**oncept |
| **POO** | **P**rogrammation **O**rientée **O**bjet |
| **RAM** | **R**andom **A**ccess **M**emory |
| **REST** | **RE**presentational **S**tate **T**ransfert |
| **RPC** | **R**emote **P**rocedure **C**all |
| **SI** | **S**ystème d'**I**nformation |
| **SLA** | **S**ervice-**L**evel **A**greement |
| **STL** | **S**tandard **T**emplate **L**ibrary |
| **UE** | **U**nreal **E**ngine |
| **VR** | **V**irtual **R**eality |
| **XR** | e**X**tended **R**eality |

# 7. DÉFINITIONS

| | |
|---|---|
| **Augmented Reality (AR)** | « La réalité augmentée, par opposition à la réalité virtuelle, conserve les objets du monde réel tels qu'ils sont et superpose des couches d'objets numériques au monde réel. Les systèmes AR intègrent trois fonctionnalités différentes qui sont 1) la combinaison du monde réel et virtuel, 2) une interaction en temps réel et 3) un enregistrement 3D précis des objets virtuels et réels. » [12] |
| **Bibliothèque logiciel** | Ensemble de programmes informatiques réutilisables dans un programme principal ou dans d'autres bibliothèques. |
| **Blueprint** | Langage de programmation informatique graphique utilisable dans l'éditeur d'Unreal Engine. |
| **C++** | Langage de programmation compilé multi paradigmes tel que la programmation procédurale, la programmation orientée objet ou la programmation générique. |
| **Extended Reality (XR)** | Le terme XR englobe les technologies de VR, AR, et MR. Là où la réalité mixte est la réunion des technologies AR et VR, la réalité étendue adresse un système qui englobe les 3 technologies (VR, AR et MR). |
| **Framework** | Un framework est une bibliothèque de développement logiciel. Un framework contient généralement de nombreuses fonctionnalités utiles à l'usage pour lequel il a été créé. Ces fonctionnalités, réutilisables, permettent au développeur de gagner du temps et de ne pas réinventer des fonctions existantes. |
| **Hardware In the Loop** | La simulation **H**ardware **I**n the **L**oop est une technique utilisée dans le développement et le test de systèmes embarqués temps réel. Il s'agit de connecter un matériel au simulateur en vue de conduire des tests de vérification, de validation ou d'intégration [14]. |
| **Métaverse** | « Le métaverse est une réalité numérique qui combine des aspects des médias sociaux, des jeux en ligne, de la réalité augmentée (AR), de la réalité virtuelle (VR) et des crypto-monnaies. Le terme métaverse fut inventé par Neal Stephenson dans son roman Snow Crash de 1992. » [15]. Dans le concept de métaverse, les mondes virtuel et réel sont en connexion permanente. Il existe simultanéité entre ces deux univers. |
| **Mixed Reality (MR)** | « La réalité mixte est la fusion des mondes réels et virtuels pour produire de nouveaux environnements et visualisations. Ici, les objets physiques et numériques coexistent et interagissent en temps réel. Contrairement à l'AR, les utilisateurs peuvent interagir avec des objets virtuels. » [12]. La réalité mixte est la réunion des technologies AR et VR. |
| **Scalabilité** | La scalabilité verticale consiste à augmenter la puissance des serveurs pour répondre au besoin de ressources alors que la demande augmente (généralement due à un nombre croissant d'utilisateur). Il s'agit d'agir sur le matériel. La scalabilité horizontale est basée sur l'ajout de serveurs en parallèle en vue de répartir la charge sur plusieurs machines (généralement, il s'agit de machines virtuelles ou de conteneurs). |
| **Tier** | Les architecture n-Tier désignent les architectures en couches. Les plus communes sont les architecture 2 et 3 tier. Dans ces dernières, le niveau 1 est la partie client, le niveau 2 les serveurs applicatifs et le niveau 3 les serveurs de base de données. |
| **Virtual Reality (VR)** | « En général, la réalité virtuelle crée un tout nouvel environnement et offre une expérience complètement immersive aux utilisateurs. Il utilise la technologie informatique pour créer une expérience simulée, qui peut être similaire ou complètement différente du monde réel. Les systèmes VR standard utilisent des casques ou des environnements multi-projections pour générer des sons et des visuels réalistes. » [12] |
| **Wrapper** | En génie logiciel, un wrapper désigne un composant qui encapsule un sous composant afin d'en offrir les fonctionnalités, sans la complexité d'utilisation (effet boite noire). |

## 8. TABLE DES FIGURES

## 9. LISTE DES TABLEAUX